\documentstyle[12pt,epsf]{article}
\newcommand{\mathbf}{{\bf}}
\newcommand{\mathrm}{{\rm}}

\begin{document}

\begin{flushright} 
USM-TH-74
\end{flushright}

\bigskip\bigskip
\centerline{\large \bf The Quark Orbital Angular Momentum 
in a Light-Cone	Representation}

\vspace{22pt}

\centerline{\bf Bo-Qiang Ma$^{a}$ and Ivan Schmidt$^{b}$ 
}

\vspace{8pt}
{\centerline {$^{a}$CCAST (World Laboratory), P.O.~Box 8730, 
Beijing 100080, China}}

{\centerline {and Institute of High Energy Physics, 
Academia Sinica,	P.~O.~Box 918(4),} 

{\centerline {Beijing 100039, China}} 

{\centerline{Email: mabq@bepc3.ihep.ac.cn}} 

{\centerline {$^{b}$Departamento de F\'\i sica, 
Universidad T\'ecnica Federico Santa Mar\'\i a,}} 

{\centerline {Casilla 110-V, 
Valpara\'\i so, Chile}}

{\centerline{Email: ischmidt@fis.utfsm.cl} }


\begin{abstract}
We perform an analysis of the quark angular momentum
in a light-cone representation by taking into account the
effect due to the Melosh-Wigner rotation
and find that there is a relativistic correction factor
connecting the quark orbital angular momentum with the
quark model spin distribution: 
$L_q(x)={<M_L(x)>}\Delta q_{QM}(x)$.
The quark orbital angular momentum $L_q(x)$ 
and the quark helicity distribution
$\Delta q(x)$ are connected to 
the quark model spin distribution
$\Delta q_{QM}(x)$ by
a relation:
$\frac{1}{2}\Delta q(x)+ L_q(x)=\frac{1}{2}\Delta q_{QM}(x)$,
which means that one can decompose the quark model spin
contribution $\Delta q_{QM}(x)$ by a quark helicity term 
$\Delta q(x)$ {\it plus} an orbital 
angular momentum term $L_q(x)$. 
There is also a new relation connecting
the quark orbital angular momentum 
with the measurable quark helicity
distribution and transversity distribution ($\delta q(x)$):
$\Delta q(x)+L_q(x)=\delta q(x)$, from which we may have new
sum rules connecting the quark orbital angular momentum 
with the nucleon axial and tensor charges.
\end{abstract}

{\centerline{PACS numbers: 
11.55.Hx, 13.60.Hb, 13.88.+e, 14.20.Dh}}


\vfill

\newpage

\section{Introduction}

The proton spin problem has received attention for about
a decade since the observation of a small value of the
integrated spin structure function for the proton in 
the experiment of polarized deep inelastic scattering (DIS)  
of leptons on the proton by the European Muon Collaboration
\cite{EMC}. The experimental results suggested a very small
quark helicity sum which is far from 1,  
the value of the quark spin 
contribution to the proton spin in the quark model. This
triggered the proton ``spin crisis" 
or ``spin puzzle" which implied
the conflict between the experimental results and the quark model,
and there have been thousands of papers related to various aspects 
on this issue \cite{Spin}. 
The prevailing viewpoint seems to be that 
the proton spin structure is in conflict with the
quark model. 

However, this viewpoint should be changed from a number of
investigations \cite{Ma91b,Bro94,Ma96,Sch97,Ma97} 
on the proton spin structure in the light-cone 
formalism, which is the most convenient framework to describe
deep inelastic scattering (DIS). 
It has been pointed out \cite{Ma91b,Bro94}
that the quark helicity ($\Delta q$) 
observed in polarized 
DIS is actually the quark spin defined 
in the light-cone formalism
and it is different from the quark spin ($\Delta q_{QM}$)
as defined in the quark model. 
Thus the small quark helicity sum
observed in polarized DIS is not 
necessarily in contradiction with
the quark model in which the proton spin 
is provided by the
valence quarks \cite{Ma96,Ma97}.
There have been studies which
support the above physical results in different
frameworks, such as in a conventional QCD Lagrangian 
based formalism \cite{Qing98}
and in a quark model approach 
in the rest reference frame \cite{Zav97}. 

In this paper we will perform 
a light-cone analysis of the quark orbital
angular momentum along with the 
developments in Refs.~\cite{Ma91b,Bro94,Sch97} 
by taking into
account the effect due 
to the Melosh-Wigner rotation \cite{MW} which
is an important ingredient 
in the light-cone formalism \cite{Bro97}. 
The light-cone formalism is suitable to describe the
relativistic many-body problem 
and there have been many successful
applications
of the light-cone quark model to various physical processes
\cite{MW2}. The effects due to the Melosh-Wigner rotation
have been calculated for 
the nucleon axial charges \cite{Ma91b,Bro94}
and magnetic moments \cite{Bro94}, and most recently for
the nucleon tensor charges \cite{Sch97}.
It will be shown that there is also 
a Melosh-Wigner rotation factor 
connecting the quark orbital angular momentum 
with the quark model spin distribution:
$L_q(x)={<M_L(x)>}\Delta q_{QM}(x)$, 
in analogy to the quark helicity
distribution \cite{Ma91b,Bro94} and transversity distribution
\cite{Sch97}.
The quark orbital angular momentum $L_q(x)$ 
and the quark helicity distribution
$\Delta q(x)$ are connected to 
the quark model spin distribution
$\Delta q_{QM}(x)$ by
a relation:
$\frac{1}{2}\Delta q(x)+ L_q(x)=\frac{1}{2}\Delta q_{QM}(x)$,
which means that one can decompose the quark model spin
contribution $\Delta q_{QM}(x)$ by a quark helicity term 
$\Delta q(x)$ {\it plus} an orbital 
angular momentum term $L_q(x)$. 
We have also a new relation connecting
the quark orbital angular momentum 
with the measurable quark helicity
distribution and transversity distribution:
$\Delta q(x)+L_q(x)=\delta q(x)$, from which we may have new
sum rules connecting the quark orbital 
angular momentum with the nucleon axial
and tensor charges.

\section{The Melosh-Wigner rotation 
in a light-cone representation}

The light-cone formalism provides a
convenient framework for 
the relativistic description of hadrons
in terms of quark and gluon degrees of freedom \cite{LCF}.
Light-cone quantization has a number of unique features
that make it appealing, most notably, the ground state
of the free theory is also a ground state of the full
theory, and the Fock expansion constructed on this
vacuum state provides a complete relativistic many-particle
basis for diagonalizing the full theory \cite{Bro94}.
As we have known, it is proper to describe deep inelastic
scattering 
as the sum of incoherent scatterings
of the incident lepton on the partons 
in the infinite momentum frame or in the light-cone formalism.
The Melosh-Wigner rotation \cite{MW} is one of the most important
ingredients of the light-cone formalism, and it relates
the light-cone (LC) spin state 
wavefunctions $q_{LC}^{\uparrow,\downarrow}$
to the ordinary instant-form (IF) spin state wavefunctions
$q_{IF}^{\uparrow,\downarrow}$ by the equation:
\begin{equation}
\begin{array}{clcr}
q_{LC}^{\uparrow}&
=&w[(k^++m)q_{IF}^{\uparrow}+k^R q_{IF}^{\downarrow}];\\
q_{LC}^{\downarrow}&
=&w[-k^L q_{IF}^{\uparrow}+(k^++m)q_{IF}^{\downarrow}],
\label{Melosh-rotation}
\end{array}
\end{equation}
where $w=[(k^++m)^2+\mathbf{k}_{\perp}^2]^{-1/2}$, 
$k^{R,L}=k^1\pm i	k^2$, and
$k^+=k^0+k^3=x {\cal M}$ in which 
the light-cone invariant mass ${\cal
M}$ of the many-body system is given 
by ${\cal M}^2=\sum_{q}\frac{m^2_q+{\mathbf
k_q}^2_{\perp}}{x_q}$. 
In a weak-binding limit one can approximate
this rotation by putting ${\cal M}=M_B$, but in our case the 
rotation applies to every individual 
spin state of a whole relativistic
many-body system, 
therefore the rotation Eq.~(\ref{Melosh-rotation}) 
should not be confused as 
a single rotation for a single free particle.

We express the instant-form 
spin state wavefunctions by the 4-dimensional
vectors $q_{IF}^{\uparrow}=(1,0,0,0)$ and 
$q_{IF}^{\downarrow}=(0,1,0,0)$, therefore we can write
the light-cone spin state wavefunctions as
\begin{equation}
\begin{array}{clcr}
q_{LC}^{\uparrow}&=&(w(k^++m),w k^R,0,0);\nonumber \\
q_{LC}^{\downarrow}&=&(-w k^L, w(k^++m),0,0).
\label{lcsw}
\end{array}
\end{equation}
Then we can write the light-cone spin space wave function of 
a bound state 
in terms of direct products of the above light-cone spin state
wavefunctions of the
individual quarks, with the many-body system keeps 
the same spin structure as in the ordinary
rest reference frame. 
In case we calculate a 
certain matrix element of a physical quantity with known operator,
we can study the effect in the spin sector by 
acting the operator directly on the corresponding
light-cone spin
state wavefunctions which is one part 
of the whole light-cone wavefunction 
for the bound system.    

Now we reanalyse the quark helicity distribution
and transversity distribution in the light-cone representation
sketched out above. The quark helicity distribution is defined
by the axial current matrix element 
\begin{equation}
\Delta q=<p,\uparrow|\overline{q} \gamma^{+} \gamma_{5}
q|p,\uparrow>.
\end{equation}
We now perform the
operator $\gamma^{+}\gamma_{5}$ 
directly on the spin space part
of the light-cone 
wavefunction for a bound system and 
the actions from the spin sector
are 
\begin{equation}
\begin{array}{clcr}
\overline{q}_{LC}^{\uparrow}
\gamma^{+}\gamma_{5}q_{LC}^{\uparrow}
&=&M_q(x,\mathbf{k}_{\perp});\\
\overline{q}_{LC}^{\downarrow}
\gamma^{+}\gamma_{5}q_{LC}^{\uparrow}
&=&0;\\
\overline{q}_{LC}^{\uparrow}
\gamma^{+}\gamma_{5}q_{LC}^{\downarrow}
&=&0;\\
\overline{q}_{LC}^{\downarrow}
\gamma^{+}\gamma_{5}q_{LC}^{\downarrow}
&=&-M_q(x,\mathbf{k}_{\perp}),
\end{array}
\end{equation}
where
\begin{equation}
M_q(x,{\mathbf k}_{\perp})
=\frac{(k^+ +m)^2-{\mathbf k}^2_{\perp}}
{(k^+ +m)^2+{\mathbf k}^2_{\perp}}
\label{eqM1} 
\end{equation}
is the Melosh-Wigner rotation factor due to the quark intrinsic
transverse motions, as derived 
in Ref.\cite{Ma91b,Bro94}. Thus we have
\begin{equation}
\Delta q (x)
=\int [{\mathrm d}^2{\mathbf k}_{\perp}] 
M_q(x,{\mathbf k}_{\perp}) 
\Delta q_{QM} (x,{\mathbf k}_{\perp})=<M_q(x)>\Delta q_{QM} (x),
\label{Melosh1}
\end{equation}
where $\Delta q_{QM}$ is the quark spin distribution in the quark
model. 

In analogy, we can perform a similar analysis of the quark
transversity distribution which is
defined by the tensor current
\begin{equation}
<p,s|\overline{q}\sigma^{\mu\nu}q|p,s>
=\delta q \overline{U}(p,s)\sigma^{\mu\nu} U(p,s),
\end{equation}
where $U(p,s)$ is the Dirac spinor of a free nucleon with momentum
$p$ and polarization vector $s$. 
Thus we can calculate the
quark transversity distribution from the $1 + $ component 
\begin{equation}
\delta q=i <p,\downarrow|\overline{q}\sigma^{1 + } q|p,\uparrow>.
\end{equation}
The actions of the operator $\sigma^{1 +}$ on the 
spin space part of the light-cone wavefunction are found to be  
\begin{equation}
\begin{array}{clcr}
\overline{q}_{LC}^{\uparrow}\sigma^{1 + } q_{LC}^{\uparrow}
&=&-2 \, w^2 k_2(k^++m);\\
\overline{q}_{LC}^{\downarrow}\sigma^{ 1+ } q_{LC}^{\uparrow}
&=&-i{\widetilde M}_q(x,{\mathbf k}_{\perp})-i\,w^2 (k1+i\,k2)^2;\\
\overline{q}_{LC}^{\uparrow}\sigma^{1+} q_{LC}^{\downarrow}
&=&i{\widetilde M}_q(x,{\mathbf k}_{\perp})+i\, w^2 (k1- i\,k2)^2;\\
\overline{q}_{LC}^{\downarrow}\sigma^{1 +} q_{LC}^{\downarrow}
&=&2 w^2 k_2 (k^++m),
\label{eqM20}
\end{array}
\end{equation}
where
\begin{equation}
{\widetilde M}_q(x,{\mathbf k}_{\perp})=\frac{(k^+ +m)^2}
{(k^+ +m)^2+{\mathbf k}^2_{\perp}}
\label{eqM2} 
\end{equation}
is the Melosh-Wigner rotation factor found in Ref.\cite{Sch97}.
One easily finds that the other additional terms
in Eq.~(\ref{eqM20}) vanish upon integration over the azimuth 
of ${\mathbf k}_{\perp}$, thus we have 
\begin{equation}
\delta q (x)
=\int [{\mathrm d}^2{\mathbf k}_{\perp}] 
{\widetilde M}_q(x,{\mathbf k}_{\perp}) 
\Delta q_{QM} (x,{\mathbf k}_{\perp})
=<{\widetilde M}_q(x)>\Delta q_{QM}(x)
\label{Melosh2}
\end{equation} 
which was first given in Ref.\cite{Sch97}. It is found that there
is a relation \cite{Sch97}: $1+M_q=2 {\widetilde M}_q$, from which
a relation connecting the quark helicity and transversity
distributions to the quark model spin distribution was suggested
\cite{Ma97}
\begin{equation}
\Delta q_{QM}(x)+\Delta q(x)= 2 \delta q(x).
\label{MSS}
\end{equation}
This relation was first proposed 
in Ref.~\cite{Ma97} and the consequences
from it's application have also been discussed \cite{Ma97,Ma97b}. 
There has been recently 
a proof of this relation in a QCD Lagrangian based
formalism \cite{Qing98}.

We proved in the above the efficiency to derive the known effects
due to the Melosh-Wigner rotation 
in the quark helicity and transversity
distributions \cite{Ma91b,Bro94,Sch97} in a simple representation 
with the light-cone spin state wavefunctions Eq.(\ref{lcsw}). 
As we have pointed out,
we can easily apply a known operator on those
spin state wavefunctions directly to study the relativistic effect
in the spin sector of the quark model. Now we make an analysis
of the quark orbital angular momentum in this light-cone
representation.  The orbital angular momentum
is defined by
\begin{equation}
{\mathbf L}_q=-i {\mathbf k} \times \nabla _{\mathbf k},
\end{equation}
and the contribution in the proton spin direction can be calculated
from the operator
\begin{equation}
{\hat L}_q=-i(k_1 \frac{\partial}{{\partial k}_2}-k_2 \frac{\partial
}{{\partial k}_1}). 
\end{equation} 
The actions of the operator ${\hat L}_q$ on the 
individual spin space part of 
the light-cone wavefunction are found to be  
\begin{equation}
\begin{array}{clcr}
\overline{q}_{LC}^{\uparrow} {\hat L}_q q_{LC}^{\uparrow}
&=&M_L(x,{\mathbf k}_{\perp});\\
\overline{q}_{LC}^{\downarrow} {\hat L}_q q_{LC}^{\uparrow}
&=&w^2(k_1+i\,k_2)(k^++m);\\
\overline{q}_{LC}^{\uparrow} {\hat L}_q q_{LC}^{\downarrow}
&=&w^2(k_1-i\,k_2)(k^++m);\\
\overline{q}_{LC}^{\downarrow} {\hat L}_q q_{LC}^{\downarrow}
&=&-M_L(x,{\mathbf k}_{\perp}),
\label{eqM3}
\end{array}
\end{equation}
where
\begin{equation}
M_L(x,{\mathbf k}_{\perp})=\frac{{\mathbf k}_{\perp}^2}
{(k^+ +m)^2+{\mathbf k}^2_{\perp}}
\label{Melosh30} 
\end{equation}
is a new Melosh-Wigner rotation factor. The other terms besides
$M_L(x,{\mathbf k}_{\perp})$ in Eq.~(\ref{eqM3}) vanish 
upon integration over the azimuth 
of ${\mathbf k}_{\perp}$, thus we have 
the quark
orbital angular momentum
\begin{equation}
L_q (x)
=\int [{\mathrm d}^2{\mathbf k}_{\perp}] M_L(x,{\mathbf k}_{\perp}) 
\Delta q_{QM} (x,{\mathbf k}_{\perp})
=< M_L(x)>\Delta q_{QM}(x)
\label{Melosh3}
\end{equation} 
which connects the quark orbital angular momentum to the quark model
spin distribution by a Melosh-Wigner rotation.

It is interesting to notice that
\begin{equation}
\frac{1}{2}M_q+M_L=\frac{1}{2},
\end{equation}
from which we have a relation
\begin{equation}
\frac{1}{2}\Delta q(x)+ L_q(x)=\frac{1}{2}\Delta q_{QM}(x),
\label{MS2}
\end{equation}
which means that we can decompose the quark model spin
distributions $\frac{1}{2}\Delta q_{QM}(x)$ into a quark helicity term
$\frac{1}{2}\Delta q(x)$ plus an orbital angular momentum
term $L_q(x)$. Thus from a relativistic viewpoint the orbital angular
momentum is non-zero even for an s-wave quark in the quark model.
This confirms the statement \cite{Ma91b} that 
the angular momentum contribution from
an s-wave quark to the proton spin should be equal to the
quark model spin $\frac{1}{2}\Delta q_{QM}$, 
but not the quark helicity $\frac{1}{2}\Delta
q(x)$. Thus the small quark helicity sum 
$\Delta \Sigma=\sum_q \Delta q(x)$ 
observed in polarized
DIS is not in conflict with the quark model. A reduction in the
quark helicity is complemented by an increase in the orbital 
angular momentum from a relativistic viewpoint. This also explains
why our result is in fact not in conflict with the conventional
statement that quark spin carries only a small part of the proton
spin if one takes $\frac{1}{2}\Delta q$ as the ``quark spin".
However, the angular momentum contribution 
from the quark to the proton spin 
should be the ``quark spin" {\it
plus} the {\it relativistic} orbital angular momentum, 
which is actually the quark model spin
$\frac{1}{2}\Delta q_{QM}$.  
From this sense, the quark orbital angular momentum 
plays an important
role in the spin content of the nucleon.
In fact, the importance of the
quark orbital angular momentum in the nucleon spin structure was
originally noticed by Sehgal \cite{Seh74}, 
and some other relevant aspects 
are also under discussion \cite{Son98}.

The above results from the Melosh-Wigner rotation are 
valid in a quite general framework of the
light-cone quark model \cite{Bro94,MW2} which is in fact
non-perturbative. 
We point out that the relation Eq.(\ref{MS2}) 
is also valid in different
approaches such as in the bag model or if one calculates the
matrix elements of the quark helicity 
and orbital angular momentum 
distributions in the nucleon rest frame with the ordinary free
Dirac spinors.
Thus this relation might be
considered as a 
{\it rather} 
model independent relation with general physical 
implications, similar to the known relation Eq.(\ref{MSS})
proposed in Ref.\cite{Ma97}. 
Even if one takes both
Eqs.(\ref{MSS}) and (\ref{MS2}) as model dependent
results, one may still make independent measurements
of the model quantity $\Delta q_{QM}(x)$ 
with either Eq.(\ref{MSS})
or Eq.(\ref{MS2}) independently, thus testing the validity
of the model results. 
We notice that the Melosh-Wigner rotation factors satisfy
a relation
\begin{equation}
M_q+M_L={\widetilde M}_q,
\end{equation}
thus 
we have 
\begin{equation}
\Delta q(x)+L_q(x)=\delta q(x),
\label{NMS}
\end{equation}
which is a new and elegant relation
directly connecting the three measurable
quantities $\Delta q(x)$, $\delta q(x)$, and $L_q(x)$.

At this point it is important to clarify the range of validity of
the particular relativistic bound-state model we are using. It 
is a three-quark valence model, which therefore does not consider
higher Fock states, and which can then be considered as a 
starting point for dynamically generated gluon and sea 
distributions \cite{gluonsea}. 
But to the extent that gluons generate the 
binding, it does contain intrinsic gluons \cite{igluon}. 
So we expect this 
model to be valid at least up to a momentum scale $Q^2 = Q_0^2$, 
where $Q_0$ can be taken as the color inverse neutral target 
size, approximately $1$ GeV$^2$. 
In practice, the light-cone quark model
has been successfully applied to processes with higher $Q^2$
of about a few  GeV$^2$ \cite{MW2,Sch}. 
Therefore it is reasonable to 
expect our relations to be approximately valid up to a  
scale $Q_0^2 \approx 1-5$ GeV$^2$ \cite{Ma97}.
Thus confirmation of the validity or invalidity of the relation
Eq.(\ref{NMS}) will be helpful 
to reveal new content concerning the spin
structure of the nucleon. 

\section{The light-cone SU(6) quark-spectator model}

We now discuss the $x$-dependent quark orbital angular 
momentum distributions $L_q(x)$ for the valence $u$ and $d$
quarks
in a light-cone SU(6) quark-spectator model
\cite{Ma96}.
The unpolarized valence quark distributions $u_v(x)$ and $d_v(x)$
are given in this model by
\begin{eqnarray} 
&&u_{v}(x)=\frac{1}{2}a_S(x)+\frac{1}{6}a_V(x);\nonumber\\
&&d_{v}(x)=\frac{1}{3}a_V(x),
\label{eq:ud}
\end{eqnarray}
where $a_D(x)$ ($D=S$ for scalar spectator or $V$ for 
axial vector
spectator) is normalized such
that $\int_0^1 {\mathrm d} x a_D(x)=3$, 
and it denotes the amplitude for quark
$q$ to be scattered while the spectator is in the diquark state $D$.
Exact SU(6) symmetry provides the relation $a_S(x)=a_V(x)$,
which implies the valence flavor symmetry $u_{v}(x)=2 d_{v}(x)$. 
This gives the prediction $F^n_2(x)/F^p_2(x)\geq 2/3$ for
all $x$,  
which is ruled out by the experimental
observation $F^n_2(x)/F^p_2(x) <  1/2$ for $x \to 1$.
The mass
difference between the scalar
and vector spectators can reproduce the $u$ and $d$ valence
quark asymmetry that
accounts for 
the observed ratio $F_2^{n}(x)/F_2^{p}(x)$ at large $x$
\cite{Ma96}. 
This
supports the quark-spectator picture of deep inelastic scattering
in which the difference between the mass of the scalar and vector
spectators is important in order to reproduce the explicit
SU(6) symmetry breaking while the bulk SU(6) symmetry of the
quark model still holds.

The quark helicity distributions
for the $u$ and $d$ quarks can be written as \cite{Ma96}
\begin{eqnarray}
&&\Delta u_{v}(x)=u_{v}^{\uparrow}(x)-u_{v}^{\downarrow}(x)
=-\frac{1}{18}a_V(x)M_q^V(x) \nonumber \\
&&\phantom{..............................................}
+\frac{1}{2}a_S(x)M_q^S(x);\nonumber\\
&&\Delta d_{v}(x)=d_{v}^{\uparrow}(x)-d_{v}^{\downarrow}(x)
=-\frac{1}{9}a_V(x)M_q^V(x),
\label{eq:sfdud}
\end{eqnarray}
in which $M_q^S(x)$ and $M_q^V(x)$ 
are the Melosh-Wigner correction factors
for the scalar and axial vector spectator-diquark cases. 
They are obtained by averaging Eq.~(\ref{eqM1})
over ${\mathbf k}_{\perp}$ with
${\cal M}^2=\frac{m^2_q+{\mathbf k}^2_{\perp}}{x}+\frac{m^2_D+{\mathbf
k}^2_{\perp}}{1-x}$, where $m_D$ is the mass of the diquark spectator,
and are unequal due to unequal spectator masses, which lead to unequal
${\mathbf k}_{\perp}$ distributions.
 From Eq.~(\ref{eq:ud}) one gets
\begin{eqnarray} 
&&a_S(x)=2u_v(x)-d_v(x);\nonumber\\
&&a_V(x)=3d_v(x).
\label{eq:qVS}
\end{eqnarray}
Combining Eqs.~(\ref{eq:sfdud}) and (\ref{eq:qVS}) we have
\begin{eqnarray} 
&&\Delta u_{v}(x)
    =[u_v(x)-\frac{1}{2}d_v(x)]M_q^S(x)-\frac{1}{6}d_v(x)M_q^V(x);
\nonumber    \\
&&\Delta d_{v}(x)=-\frac{1}{3}d_v(x)M_q^V(x).
\label{eq:dud}
\end{eqnarray}
Thus we arrive at simple relations \cite{Ma96} between the polarized
and unpolarized quark distributions for the valence $u$ and $d$
quarks. 
The relations (\ref{eq:dud})
can be considered as the results of the conventional
SU(6) quark model, and which 
explicitly take into account the Melosh-Wigner rotation effect
\cite{Ma91b,Bro94}
and the flavor asymmetry introduced by the
mass difference between the scalar and vector
spectators \cite{Ma96}.

The extension of relations Eq.~(\ref{eq:dud}) to the
quark orbital angular momentum distributions 
$L_q(x)$ is straightforward: 
we can simply replace $M_q^S(x)$ and $M_q^V(x)$
by $M_L^S(x)$ and $M_L^V(x)$,
\begin{eqnarray} 
&&L_u(x)
    =[u_v(x)-\frac{1}{2}d_v(x)]M_L^S(x)
-\frac{1}{6}d_v(x)L_L^V(x); \nonumber \\
&&L_d(x)=-\frac{1}{3}d_v(x)M_L^V(x).
\label{eq:dudT}
\end{eqnarray}
The $x$-dependent Melosh-Wigner rotation factors
$M_q^S(x)$ and $M_q^V(x)$ have been calculated \cite{Ma96} 
and an asymmetry between $M_q^S(x)$ 
and $M_q^V(x)$ was found. Since
$\frac{1}{2}M_q^S(x)+M_L^S(x)=\frac{1}{2}$ and
$\frac{1}{2}M_q^V(x)+M_L^V(x)=\frac{1}{2}$, there should be also
an asymmetry between $M_L^S(x)$ and $M_L^V(x)$ in an opposite
direction. The corresponding asymmetry between the Melosh-Wigner
rotation factors with the scalar and vector spectators will
produce a flavor asymmetry between the averaged Melosh-Wigner
rotation factor for the $u$ and $d$ quarks.
Contrary to the quark helicity case where the
asymmetry causes a reduction of the quark helicity sum compared to
the flavor symmetric case, the flavor asymmetry of the Melosh-Wigner
rotation factors in the
quark orbital angular momentum case produces an increase of the
quark orbital angular momentum sum compared to the
flavor symmetric case.

The calculated polarization asymmetries
$A_1^N=2 x g_1^N(x)/F_2^N(x)$, 
including the Melosh-Wigner rotation,
have been found \cite{Ma96} to be in reasonable agreement
with the experimental data, at least for $x \geq 0.1$.
A large asymmetry between $M_q^S(x)$ and $M_q^V(x)$
leads to a better fit to the data than that 
obtained
from a small asymmetry. 
Therefore it is
reasonable to expect that the 
calculated $L_q(x)$  
may lead to predictions close to the real situation.
In Fig.~(\ref{ms2f1}) 
we present the calculated $\Delta q(x)$ and $L_q(x)$
for the $u$ and $d$ valence quarks.

\vspace{0.5cm}
\begin{figure}[htb]
\begin{center}
\leavevmode {\epsfysize=10cm \epsffile{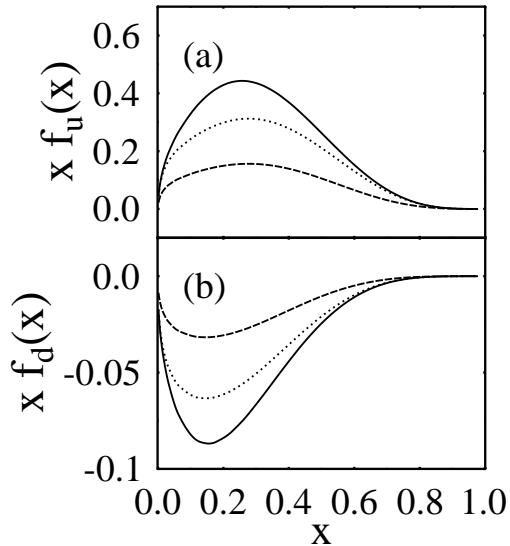}} 
\end{center}
\caption[*]{\baselineskip 13pt 
The $x$-dependent quark spin distributions $x \Delta q_{QM}(x)$ (solid
curves),
helicity distributions $x \Delta q(x)$ (dotted curves), 
and orbital angular momentum distributions 
$x L_q(x)$ (dashed curves) in the light-cone
SU(6) quark-spectator model by using
Eqs.~(\ref{eq:dud}-\ref{eq:dudT}), 
with the Gl\"uck-Reya-Vogt parameterization 
\cite{GRV95} 
of unpolarized quark distributions as input:
(a) for $u$ quarks; (b) for $d$ quarks. }
\label{ms2f1}
\end{figure}

\section{New sum rules for the quark orbital angular momentum}

As we have known, parton sum rules have played important roles
in the understanding of the quark-gluon structure of the nucleons.
For example, the Gottfried sum rule \cite{GSR} 
was proposed initially
for the hope to destroy the quark model \cite{GSR2}. 
However, the confirmation
of the roughly validity of the Gottfried sum rule in the early
deep inelastic experiments was important for identifying the quantum
numbers of partons with those of quarks, thus confirmed the quark
model. The recent observation of the Gottfried sum rule violation
in the refined measurements \cite{NMC91} 
revealed the flavor asymmetry between
the $u$ and $d$ sea quarks \cite{Pre91}. 
Thus the validity or invalidity of a 
parton sum rule is useful in order to reveal 
possible new physics.

New sum rules for the quark tensor charges 
have been recently proposed
\cite{Ma97b} based on the relation Eq.(\ref{MSS}) \cite{Ma97}. 
Based on the fact that the
anti-quarks in the nucleon sea are likely 
to be non-polarized from theoretical
considerations \cite{Che95,Bro96} and experimental evidence
\cite{NSMCN}, 
a Bjorken-like sum rule for the isovector tensor charge is found
\begin{equation}
\delta U -\delta D
=\frac{1}{2}(\frac{g_A}{g_V}+\frac{5}{3} c_1)
\label{sum1}
\end{equation}
where $g_A/g_V$ might be the value from the neutron $\beta$
decay or $g_A/g_V=6 (\Gamma^p-\Gamma^n)$ (where 
the integrals
$\Gamma^p=\int_0^1 {\mathrm d} x g_1^p(x)$
and $\Gamma^n=\int_0^1 {\mathrm d} x g_1^n(x)$ have been measured
from 
polarized DIS experiments), 
and $c_1$ is an unknown correction factor reflecting the deviation
from the naive quark model value
$\Delta U_{QM}-\Delta D_{QM}=5/3$. 
Similarly, there is also a sum rule for the isoscalar tensor charge
\begin{equation}
\delta U +\delta D=
\frac{9}{5}(\Gamma^p+\Gamma^n)
-\frac{1}{5}\Delta S+\frac{1}{2} c_2,
\label{sum2}
\end{equation}
where $\Delta S$ is the quark axial charge for the strange
flavor and $c_2$ is another unknown correction factor
reflecting the deviation
from the naive quark model value
$\Delta U_{QM}+\Delta D_{QM}=1$. In fact, 
$c_2$ can be simply taken as the fraction
of the proton spin carried by the $u$ and $d$ quarks in the quark
model picture.

 From the same consideration, we may apply the relation 
Eq.(\ref{MS2}) and arrive at new sum rules for
the quark orbital angular momentum $L_q=\int_0^1 {\mathrm d} x
L_q(x)$. Corresponding to
Eq.(\ref{sum1}), we have, 
for the isovector quark orbital angular momentum,
\begin{equation}
L_u -L_d
=\frac{1}{2}(-\frac{g_A}{g_V}+\frac{5}{3} c_1);
\label{nsum1}
\end{equation}
and corresponding to
Eq.(\ref{sum2}), we have, 
for the isoscalar quark orbital angular momentum,
\begin{equation}
L_u +L_d 
=-\frac{9}{5}(\Gamma^p+\Gamma^n)
+\frac{1}{5}\Delta S+\frac{1}{2} c_2.
\label{nsum2}
\end{equation}
We notice that above two sum rules do not require the non-polarized
anti-quarks which are required for the validity of the sum rules 
Eqs.(\ref{sum1}) and (\ref{sum2}). 

 From Eqs.~(\ref{nsum1}) and (\ref{nsum2}),
we can predict the quark orbital angular momenta $L_u-L_d$
and $L_u+L_d$ by use of 
the measurable quantities $\Gamma_p$, $\Gamma_n$,
$g_A/g_V$ and $\Delta S$, and the correction factors $c_1$ and
$c_2$ with limited uncertainties.
The quantities $\Gamma^p$ and $\Gamma^n$ at several
different $Q^2$ have been
measured from polarized DIS experiments \cite{EMC,SMC,E142,E143}, and 
$\Delta S$ has also been
extracted from analysis of the polarized DIS data
and it might range from about -0.01 \cite{Bro96}
to -0.13 \cite{Ell95b}.
The value of $\Delta S$ from those analysis is sensitive
to the assumption of SU(3) symmetry. It would be
better to measure $\Delta S$
from other independent processes and there have been suggestions
for this purpose \cite{Lu95,Ell95}.
Nevertheless, we notice that the predicted
values of $L_u$ and $L_d$ are not sensitive to
$\Delta S$.
The quantities $c_1$ and $c_2$ 
reflect the deviation
from the naive quark model and their uncertainties can be 
estimated from known theoretical considerations. For example,
from the lattice QCD results
for the axial charge 
$\Delta \Sigma =0.18 \pm 0.10$ \cite{lQCD1} and
the tensor charge $\delta \Sigma =0.562 \pm 0.088$ \cite{lQCD2}, we
found
$\sum \Delta q_{QM}= 0.94 \pm 0.28$
from the relation
Eq.~(\ref{MSS}) and 
this supports
the naive quark picture that the spin of the proton is
mostly carried by quarks \cite{Ma97}. 
Thus $c_1$ and $c_2$ should be close
to the above value for $\sum \Delta q_{QM}$ from a reasonable
consideration. Considering that the gluon may also play an important
role in the spin structure of 
the nucleon \cite{Gluon}, we adopt a large
uncertainty $c_2=0.5 \to  1$ (which corresponds to a gluon spin
contribution in the range 50\% to 0\%) and $c_1=0.7 \to 1$ in 
comparison of $c_2=0.75 \to 1$ and $c_1=0.9 \to 1$ adopted in 
\cite{Ma97}. In case $g_A/g_V=6 (\Gamma^p-\Gamma^n)$ is adopted,
for $\Gamma^p({\rm E143})=0.127$ and
$\Gamma^n({\rm E143})=-0.037$ at $\langle Q^2 \rangle=3$ GeV$^2$
\cite{E143},
we have 
\begin{equation}
\begin{array}{clcr}
L_u-L_d= 0.09 \to 0.34;\\
L_u+L_d= 0.09 \to 0.34,
\label{con1}
\end{array}
\end{equation}
and for $\Gamma^p({\mathrm SMC})=0.136$ and
$\Gamma^n({\mathrm SMC})=-0.063$ at $\langle Q^2 \rangle=10$ GeV$^2$
\cite{SMC},
we have
\begin{equation}
\begin{array}{clcr}
L_u-L_d=-0.01 \to 0.24 ;\\
L_u+L_d=0.12  \to 0.37.
\label{con2}
\end{array}
\end{equation}
In case the value $g_A/g_V=1.2573$ from neutron $\beta$ decay is
adopted, 
for the isovector
component Eq.(\ref{nsum1}), 
we obtain
\begin{equation}
L_u-L_d= -0.05 \to 0.20.
\label{con1b}
\end{equation}
Combining the results in Eqs.(\ref{con1}), (\ref{con2}) and
(\ref{con1b}),
and taking into account also the uncertainties 0.05 introduced by the
data,
we obtain
\begin{equation}
\begin{array}{clcr}
L_u-L_d= -0.10 \to 0.39;\\
L_u+L_d= 0.04  \to 0.42.
\end{array}
\end{equation}
Progress in the precision of the data and
in the knowledge of the correction factors can further
constrain the results.

In fact, we can use the two sets of sum rules, Eqs.(\ref{sum1})
and (\ref{sum2}) {\it and} Eqs.(\ref{nsum1}) and (\ref{nsum2}),
as two independent measures of the quantities $c_1$ and $c_2$
which might be considered as model quantities with no universal
significance. From one set 
of above sum rules one can measure the quantities $c_1$ and $c_2$, 
and then use them as inputs to another set of sum rules
to make the prediction. However, by use of the relation
Eq.(\ref{NMS}) we may have a new sum rule which directly connects
the quark orbital angular momentum to the quark axial and tensor
charges
\begin{equation}
L_q=\delta Q -\Delta q,
\label{nsum}
\end{equation}
which is much simple. From the lattice QCD calculations
we have the axial charge 
$\Delta \Sigma =0.18 \pm 0.10$ \cite{lQCD1} and
the tensor charge $\delta \Sigma =0.562 \pm 0.088$ \cite{lQCD2},
thus the total quark orbital angular momentum from the new
sum rule Eq.(\ref{nsum}) should be
\begin{equation}
\Sigma L_q=\delta \Sigma - \Delta \Sigma=0.38 \pm 0.13,
\end{equation} 
which implies that $76 \pm 26$\% of the proton spin
is due to the
quark {\it relativistic} orbital motions from the known lattice
results for the nucleon axial and tensor charges.

In similar to the relation $\Delta q_{QM}(x)+\Delta q(x)=2 \delta
(x)$ discussed in Ref.\cite{Ma97,Ma97b}, the relations 
$\frac{1}{2}\Delta q(x)+L_q(x)=\frac{1}{2}\Delta q_{QM}(x)$
and $\delta q(x)+L_q(x)=\Delta q(x)$ are valid for each
flavor quark, likewise for each flavor anti-quark. The sum rule
Eq.(\ref{nsum}) will require non-polarized anti-quarks
due to the charge conjugation properties of the tensor current.
However, as have been pointed out previously \cite{Ma97,Ma97b}, 
the non-polarization
of anti-quarks has been clearly predicted in a
broken-U(3) version of the chiral quark model \cite{Che95}
and in the meson-baryon fluctuation model \cite{Bro96}.
There has been an explicit measurement of the helicity distributions
for the individual $u$ and $d$ valence and sea quarks by the Spin
Muon Collaboration (SMC) \cite{NSMCN}. The refined SMC measurement 
of the helicity distributions
for the $u$ and $d$ anti-quarks are consistent with zero in agreement
with the above predictions.
Therefore the small anti-quark polarization 
can be considered as a fact supported by both experimental 
measurement \cite{NSMCN} and theoretical
considerations \cite{Che95,Bro96}. We need to indicate that we have
neglected the contributions from anti-quarks, gluons, $Q^2$
dependence due to higher twist effects, and different evolution
behaviors of quark distributions {\it et al.} in the above
analysis. In principle the corrections due to 
these sources can be further taken into account from the 
theoretical and experimental progress and they should be topics
for further study.  

\section{Summary}

In summary,
we performed an analysis of the quark angular momentum
in a light-cone representation by taking into account the
effect due to the Melosh-Wigner rotation
and found a similar relativistic correction factor
connecting the quark orbital angular momentum to the
quark model spin distribution: $L_q(x)=<M_L(x)>\Delta q_{QM}(x)$.
The quark orbital angular momentum $L_q(x)$ 
and the quark helicity distribution
$\Delta q(x)$ are connected to the quark model spin distribution
$\Delta q_{QM}(x)$ by
a relation:
$\frac{1}{2}\Delta q(x)+ L_q(x)=\frac{1}{2}\Delta q_{QM}(x)$,
which means that one can decompose the quark model spin
contribution $\Delta q_{QM}(x)$ by a quark helicity term 
$\Delta q(x)$ {\it plus} an orbital 
angular momentum term $L_q(x)$. 
There is also a new relation connecting
the quark orbital angular momentum with the measurable quark helicity
distribution and transversity distribution:
$\Delta q(x)+L_q(x)=\delta q(x)$, from which we have a new
sum rule connecting the quark orbital angular momentum 
with the nucleon axial
and tensor charges. Two other new sum rules for the quark orbital 
angular momentum are proposed and the values for the
isovector and isoscalar quark orbital angular momenta 
$L_u-L_d$ and $L_u+L_d$ are estimated
from the measured quantities $\Gamma^p$, $\Gamma^n$,
$g_A/g_V$ and
$\Delta S$, and two model correction factors with limited
uncertainties.
We also calculated the $x$-dependent quark orbital angular momentum
distributions
for the $u$ and $d$ valence quarks in a light-cone 
SU(6) quark-spectator model.

\bigskip
{\bf Acknowledgments: } 
We would like to thank S.J.~Brodsky
and J.~Soffer for helpful discussions. This work is 
partially supported by National Natural 
Science Foundation of China 
under Grant No.~19605006, Fondecyt (Chile) under grant 1960536, 
by a C\'atedra Presidencial (Chile),
and by Fundaci\'on Andes (Chile).


\newpage
\begin{thebibliography}{99}

\bibitem{EMC}
                EM Collab., J.~Ashman {\it et al.},
                Phys.~Lett.~{\bf B 206}, 364 (1988);
                Nucl.~Phys.~{\bf B 328}, 1 (1989).

\bibitem{Spin}
For reviews, see, e.g., 

H.-Y.~Cheng, Int.~J.~Mod.~Phys.~{\bf A 11}, 5109 (1996);

G.P.~Ramsey, Prog.~Part.~Nucl.~Phys.~{\bf 39}, 599 (1997).


\bibitem{Ma91b}
B.-Q.~Ma, J. Phys. {\bf G 17}, L53 (1991);

B.-Q.~Ma and Q.-R.~Zhang,
Z.~Phys. {\bf C 58}, 479 (1993).

\bibitem{Bro94}

S.J.~Brodsky and F.~Schlumpf, Phys.
Lett. {\bf B 329}, 111 (1994).

\bibitem{Ma96}
B.-Q.~Ma, Phys. Lett. {\bf B 375}, 320 (1996);

B.-Q.~Ma and A.~Sch\"afer, Phys. Lett. {\bf B 378}, 307 (1996).

\bibitem{Sch97} I.~Schmidt and J.~Soffer, Phys.~Lett.~{\bf B 407},
331 (1997).

\bibitem{Ma97}
B.-Q.~Ma, I.~Schmidt, and J.~Soffer, hep-ph/9710247,
to be published in Phys. Lett. {\bf B}.

\bibitem{Qing98} D.~Qing, X.-S.~Chen, and F.~Wang, hep-ph/9802425.

\bibitem{Zav97} P.~Zavada, Phys.~Rev. {\bf D 56}, 5834 (1997);
hep-ph/9803443.

\bibitem{MW}
H.J.~Melosh, Phys.~Rev. {\bf D 9}, 1095 (1974); 

E.~Wigner, Ann. Math. {\bf 40}, 149 (1939).

\bibitem{Bro97}

See, e.g., S.J.~Brodsky, H-C.~Pauli, and S.S.~Pinsky,
Phys. Rep. {\bf 301}, 299 (1998), 
and references therein.

\bibitem{MW2}
See, e.g., 

L.A.~Kondratyuk and M.V.~Terent'ev, Yad.Fiz {\bf 31}, 1087 (1980) [
Sov. J. Nucl. Phys. {\bf 31}, 561 (1980)];

P.L.~Chung, F.~Coester, B.D.~Keister, and W.N.~Polyzou,
Phys.~Rev. {\bf C 37}, 2000 (1988);

H.J.~Weber, Ann.~Phys.~(N.Y.) {\bf 207}, 417 (1991);

W.~Jaus, Phys.~Rev. {\bf D 41}, 3394 (1990); {\bf D 44}, 2851 (1991);

P.L.~Chung and F.~Coester, Phys. Rev. {\bf D 44}, 229 (1991);

B.-Q.Ma, Z.~Phys. {\bf A 345}, 325 (1993);

F.~Schlumpf, Phys.~Rev. {\bf D 48}, 4478 (1993);

F.~Schlumpf and S.J.~Brodsky, Phys.~Lett. {\bf 360}, 1 (1995).

T.~Huang, B.-Q.Ma, and Q.-X.~Shen, Phys.~Rev. {\bf D 49}, 1490 (1994);

B.-Q.~Ma and T.~Huang, J.~Phys. {\bf G 21}, 765 (1995); 

F.-G.~Cao, J.~Cao, T.~Huang, and B.-Q.~Ma, Phys.~Rev. {\bf D 55},
7107 (1997).

\bibitem{LCF}
See, {e.g.}, 
S. J. Brodsky and G. P. Lepage, 
in {\it Perturbative Quantum  Chromodynamics}, edited by
A. H. Mueller (Singapore, World Scientific, 1989), p. 93;

S. J. Brodsky, T. Huang, and G. P. Lepage, in {\it Particles and
Fields-2}, Proceedings of the Banff Summer Institute, Banff, Alberta,
1981, edited by A. Z. Capri and A. N. Kamal 
(Plenum, New York,1983), p. 143;
G. P. Lepage, S. J. Brodsky, 
T. Huang, and P. B. Mackenize, {\it ibid.},
p. 83.

\bibitem{LCS}
G.P.~Lepage and S.J.~Brodsky, Phys.~Rev.~{\bf D 22}, 2157 (1980);

W.~Konen and H.J.~Weber, Phys.~Rev. {\bf D 41}, 2201 (1990).

\bibitem{Ma97b} B.-Q.~Ma and I.~Schmidt, hep-ph/9711326,
to be published in J. Phys. {\bf G}.

\bibitem{Seh74}
L.M.~Sehgal, Phys.~Rev.~{\bf D 10}, 1663 (1974).

\bibitem{Son98}
See, e.g., X.~Song, hep-ph/9801332; 
P.~Hoodbhoy, X.~Ji, and W.~Lu, hep-ph/9804337; 
and references therein.

\bibitem{gluonsea}
F.~Martin, Phys.~Rev.~{\bf D 19}, 1382 (1979);

M.~Gl\"uck, E.~Reya, and W.~Vogelsang, 
Nucl.~Phys.~{\bf B 329}, 347 (1990). 

\bibitem{igluon}

S.J.~Brodsky and I.~Schmidt, Phys.~Lett.~{\bf B 234}, 144 (1990); 

P.~Hoyer and D.P.~Roy, Phys.~Lett.~{\bf 410}, 63 (1997).

\bibitem{Sch}
F.~Schlumpf, J.~Phys.~{\bf G 20}, 237 (1994);

See also, F.Schlumpf, Ph.D. thesis, 
University of Zurich, 1992. 

\bibitem{GRV95}
               M.~Gl\"uck, E.~Reya, and A.~Vogt,
               Z.~Phys.~{\bf C 67}, 433 (1995). We use the LO set
               at $Q^2=5$ GeV$^2$ in the calculations. 

\bibitem{GSR}   K.~Gottfried,
                Phys.~Rev.~Lett. {\bf 18}, 1174 (1967).

\bibitem{GSR2}
Proc. 1967 International Symposium on Electron and Photon 
Interactions at High Energy, Stanford, California, September 5-9,
167.

\bibitem{NMC91} NM Collab., P.~Amaudruz {\it et al.},
                Phys.~Rev.~Lett. {\bf 66}, 2712 (1991);

                M.~Arneodo {\it et al.},
                Phys.~Rev.~{\bf D 50}, R1 (1994).

\bibitem{Pre91}
G.~Preparata, P.G.~Ratcliffe, and J.~Soffer,
Phys.~Rev.~Lett. {\bf 66}, 687 (1991).

\bibitem{Che95}
T.P.~Cheng and L.F.~Li, 
Phys.~Lett. {\bf B 366}, 365 (1996).

\bibitem{Bro96}
S.J.~Brodsky and B.-Q.~Ma, Phys.~Lett.~{\bf B 381}, 317 (1996).

\bibitem{NSMCN}
SM Collab., B.~Adeva {\it et al.}, Phys.~Lett.~{\bf B 369}, 93
(1996);  for most recent refined results, see, 
Phys.~Lett.~{\bf 420}, 180 (1998). 
\bibitem{SMC}
                SM Collab., B.~Adeva {\it et al.},
                Phys.~Lett.~{\bf B 302}, 533 (1993);
                {\bf B 357}, 248 (1995);
                D.~Adams {\it et al.}, {\it ibid.}
                {\bf B 329}, 399 (1994); {\bf B 339}, 332(E) (1994).

\bibitem{E142}
                E142 Collab., P.L.~Anthony {\it et al.},
                Phys.~Rev.~Lett.~{\bf 71}, 959 (1993).

\bibitem{E143}
                E143 Collab., P.L.~Anthony {\it et al.},
                Phys.~Rev.~Lett.~{\bf 74}, 346 (1995);
                K.~Abe {\it et al.}, {\it ibid.} {\bf 75}, 25
                (1995).

\bibitem{Ell95b}
J.~Ellis and M.~Karliner, Phys.~Lett.~{\bf B 341}, 397 (1995).

\bibitem{Lu95}
W.~Lu and B.~-Q.~Ma, Phys.~Lett.~{\bf B 357}, 419 (1995);

W.~Lu, Phys.~Lett.~{\bf B 373}, 223 (1996).

\bibitem{Ell95}
J.~Ellis, D.~Kharzeev, and A.~Kotzinian, Z.~Phys.~{\bf C 69}, 467
(1995).

\bibitem{lQCD1}
M.~Fukugita, Y.~Kuramashi, M.~Okawa, and A.~Ukawa,
Phys.~Rev.~Lett.~{\bf 75}, 2092 (1995).

\bibitem{lQCD2}
S.~Aoki, M.~Doui, T.~Hatsuda, and Y.~Kuramashi, 
Phys.~Rev.~{\bf D 56}, 433 (1997).

\bibitem{Gluon}

See, e.g., G.~Altarelli and G.G.~Ross, Phys.~Lett.~{\bf B 212}, 391
(1988);

R.D.~Carlitz, J.C.~Collins, and A.H.~Mueller, Phys.~Lett.~{\bf 214},
229 (1988);

A.V.~Efremov, J.~Soffer, and O.V.~Teryaev,
Nucl.~Phys. {\bf B 346}, 97 (1990).

For most recent discussions, see, e.g., J.~Soffer and
O.V.~Teryaev, 
Phys. Lett.	{\bf 419}, 400 (1998); V.~Barone, T.~Calarco, and
A.~Drago, hep-ph/9801281; and references therein.

\nonfrenchspacing
\end{thebibliography}
\end{document}